\documentclass{iau} 
\usepackage{graphicx}
\usepackage{natbib}

\pubyear{2022}
\volume{372}
\jname{The Era of Multi-Messenger Solar Physics}
\editors{G.~Cauzzi \& A.~Tritschler, eds.}

\title[Active region NOAA 12673 and X9.3 flare of 6 September 2017] %% give here short title %%
{Origin of extreme solar eruptive activity from the active region NOAA 12673 and the largest flare of solar cycle 24}

\author[Joshi \& Mitra]   %% give here short author list %%
{Bhuwan Joshi \and Prabir K. Mitra}

\affiliation{Udaipur Solar Observatory, Physical Research Laboratory, Udaipur 313 001, India \\ [\affilskip]}

\begin{document}

\maketitle

\begin{abstract}
During 2017, when the Sun was moving toward the minimum phase of solar cycle 24, an exceptionally eruptive active region (AR) NOAA 12673 emerged on the Sun during August 28-September 10. During the highest activity level, the AR turned into a $\delta$-type sunspot region, which manifests the most complex configuration of magnetic fields from the photosphere to the coronal heights. The AR 12673 produced four X-class and 27 M-class flares, along with numerous C-class flares, making it one of the most powerful ARs of solar cycle 24. Notably, it produced the largest flare of solar cycle 24, namely, the X9.3 event on 2017 September 6. In this work, we highlight the results of our comprehensive analysis involving multi-wavelength imaging and coronal magnetic field modeling to understand the evolution and eruptivity from AR 12673. We especially focus on the morphological, spectral and kinematical evolution of the two X-class flares on 6 September 2017. We explore various large- and small-scale magnetic field structures of the active region which are associated with the triggering and subsequent outbursts during the powerful solar transients. 

\end{abstract}

%\firstsection % if your document starts with a section, you can
              
\section{Introduction}

NOAA 12673 was one of the most flare productive solar ARs of cycle 24. It
underwent a rapid evolution in magnetic complexity which resulted in tremendous flare productivity. Multi-instrument observations of NOAA 12673, coupled with modeling analysis of coronal magnetic fields, have provided us with a unique
opportunity to investigate various physical parameters that play vital role in the onset of large flares and influence the early CME dynamics. Notably, NOAA
12673 produced the largest flare of the cycle, an X9.3 flare on 6 September 2017. 

In view of its uniqueness in the flare productivity, naturally the active region has already been subjected to a number of studies that involve different aspects of the long-term evolution of the AR as well as flare case studies  \citep[e.g.,][etc.]{Yang2017, Gary2018, Hou2018, Liu2019, Mitra2018, Mitra2020, Romano2018, Seaton2018, Verma2018, Veronig2018, Moraitis2019, Chen2020}.

In this article, we investigate the evolution and magnetic characteristics
of NOAA 12673. We briefly present the multi-wavelength imaging of the homologous X-class flares of intensities X2.2 and X9.3 occurred on 6 September 2017 to understand the triggering mechanism and stages of energy release. The complexity of photospheric and coronal magnetic fields prior to the X-class flares are explored to assess the link between the large- and small-scale magnetic field structures of the solar active region.

\begin{table}[t!]
\centering
\caption{Summary of the large flares (GOES class M1 or above) originated from the active region NOAA 12673}
\vskip 0.05 in
%\begin{ruledtabular}

\label{table1}
\begin{tabular}{cccccc}
\hline
\hline
Flare&Date&AR conf.&Location&Flare timing (UT)&GOES\\
Id.&(Sep '17)&&&Start/ Peak/ End&class\\
\hline
F$_1$&4&$\beta\gamma$&S10W04&05:36/ 05:49/ 06:05&M1.2\\
F$_2$&&&S10W08&15:11/ 15:30/ 15:33&M1.5\\
F$_3$&&&S07W11&18:05/ 18:22/ 18:31&M1.0\\
F$_4$&&&S09W11&18:46/ 19:37/ 19:52&M1.7\\
F$_5$&&&S10W11&19:59/ 20:02/ 20:06&M1.5\\
F$_6$&&&S10W11&20:28/ 20:33/ 20:37&M5.5\\
F$_7$&&&S09W12&22:10/ 22:14/ 22:19&M1.7\\
F$_8$&5&$\beta\gamma\delta$&S09W14&01:03/ 01:08/ 01:11&M4.2\\
F$_9$&&&S09W15&03:42/ 03:51/ 04:04&M1.0\\
F$_{10}$&&&S11W18&04:33/ 04:53/ 05:07&M3.2\\
F$_{11}$&&&S11W19&06:33/ 06:40/ 06:43&M3.8\\
F$_{12}$&&&S10W23&17:37/ 17:43/ 17:51&M1.5\\
F$_{13}$&6&$\beta\gamma\delta$&S08W32&08:57/ 09:10/ 09:17&X2.2\\
F$_{14}$&&&S09W34&11:53/ 12:02/ 12:10&X9.3\\
F$_{15}$&&&S08W36&15:51/ 15:56/ 16:03&M2.5\\
F$_{16}$&&&S08W38&19:21/ 19:30/ 19:35&M1.4\\
F$_{17}$&&&S08W40&23:33/ 23:39/ 23:44&M1.2\\
F$_{18}$&7&$\beta\gamma\delta$&S08W44&04:59/ 05:02/ 05:08&M2.4\\
F$_{19}$&&&S07W46&09:49/ 09:54/ 09:58&M1.4\\
F$_{20}$&&&S07W46&10:11/ 10:15/ 10:18&M7.3\\
F$_{21}$&&&S08W48&14:20/ 14:36/ 14:55&X1.3\\
F$_{22}$&&&S11W54&23:50/ 23:59/ 00:14 (8 Sep) &M3.9\\
F$_{23}$&8&$\beta\gamma\delta$&S09W55&02:19/ 02:24/ 02:29&M1.3\\
F$_{24}$&&&S07W55&03:39/ 03:43/ 03:45&M1.2\\
F$_{25}$&&&S09W57&07:40/ 07:49/ 07:58&M8.1\\
F$_{26}$&&&S09W63&15:09/ 15:47/ 16:04&M2.9\\
F$_{27}$&&&S08W69&23:33/ 23:44/ 23:56&M2.1\\
F$_{28}$&9&$\beta\gamma\delta$&S11W70&04:14/ 04:28/ 04:43&M1.1\\
F$_{29}$&&&S08W74&11:50/ 11:04/ 11:42&M3.7\\
F$_{30}$&&&S09W88&22:04/ 23:53/ 01:30 (10 Sep) &M1.2\\
F$_{31}$&10&$\beta\gamma\delta$&S08W88&15:35/ 16:06/ 16:31&X8.2\\
\hline
\end{tabular}
%\end{ruledtabular}
\end{table}

\begin{figure}[h]
\begin{center}
\includegraphics[width=\textwidth]{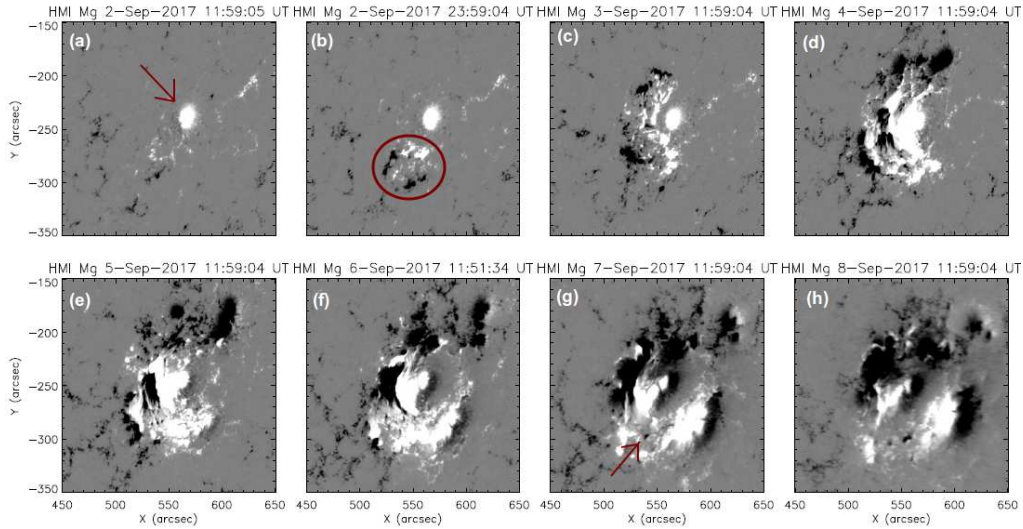} 
\caption{ Series of HMI continuum images showing the build up and evolution
of the active region NOAA 12673. All the images were rotated to the co-ordinate
corresponding to 6 September 2017 12:00 UT. The arrow in panel (a) indicate the
positive polarity sunspot. The oval shape in panel (b) indicates the initial flux emergence on 3 September 2017. The arrow in panel (g) indicates emergence of negative flux regions within the positive polarity regions during the decay phase of the active region.} \label{fig:magnetogram}

\end{center}
\end{figure}

\section{NOAA 12673: Morphology, evolution, and eruptivity}

\subsection{Evolution and activity}
The active region NOAA 12673 appeared on the eastern limb of the Sun as a simple unipolar sunspot. Subsequently, the active region exhibited the constant emergence of magnetic flux of both the polarities as well as photospheric motions including shearing and rotational motions (Figure~\ref{fig:magnetogram}). This led to the quick evolution of the region into the most complex configuration, namely the $\delta$-type, resulting a series of large eruptive flares. Notably, the AR produced 4 X-class and 27 M-class flares along with numerous smaller events. In Table~\ref{table1}, a summary of the flares of GOES class M and above is given. From Table~\ref{table1}, we note that the first M class flare from NOAA 12673 originated on 4 September 2017 following which 12 further M-class flares occurred within 2 days. After producing 2 homologous X-class flares on 6 September 2017 (see Section~\ref{sec2}), the AR produced another X-class flare (X1.3) on the next day. Subsequently, on 10 September, a large X8.2 flare occurred while the AR was transiting toward the far side of the Sun.

\begin{figure}[h]
\begin{center}
\includegraphics[width=0.9\textwidth]{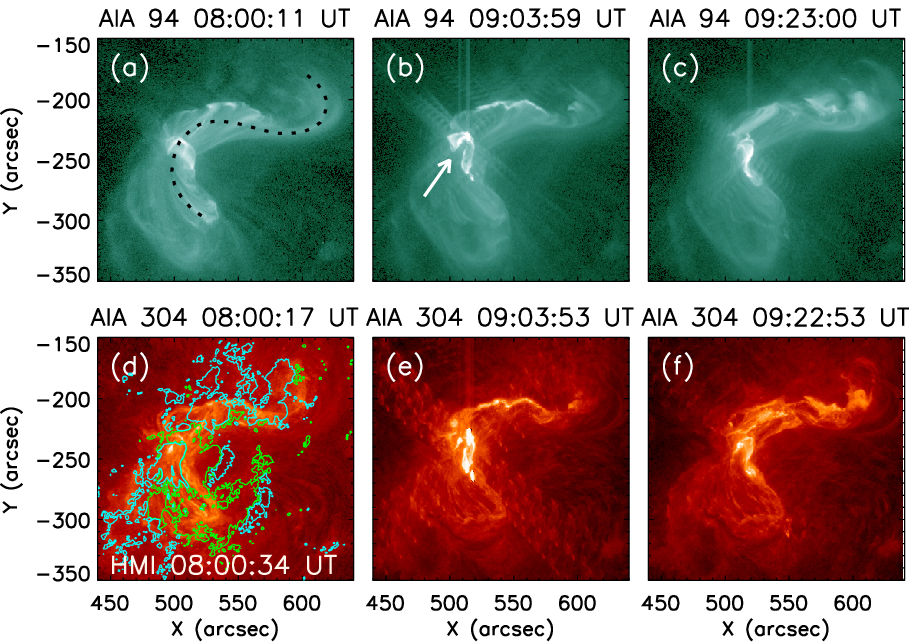} 
\caption{AIA 94 \AA~(panels (a)-(c)) and 304 \AA~(panels (d)-(f)) showing the evolution of the X2.2 flare on 6 September 2017. The dashed curve in panel (a) highlights the pre-flare sigmoidal structure. The arrow in panel (b) indicates a hot channel. Co-temporal HMI LOS magnetogram contours are plotted over panel (d). The green and sky-blue contours refer to positive and negative field, respectively. Contour levels are $\pm$150 G.} \label{fig:X2flare}

\end{center}
\end{figure}

\begin{figure}[h]
\begin{center}
\includegraphics[width=0.9\textwidth]{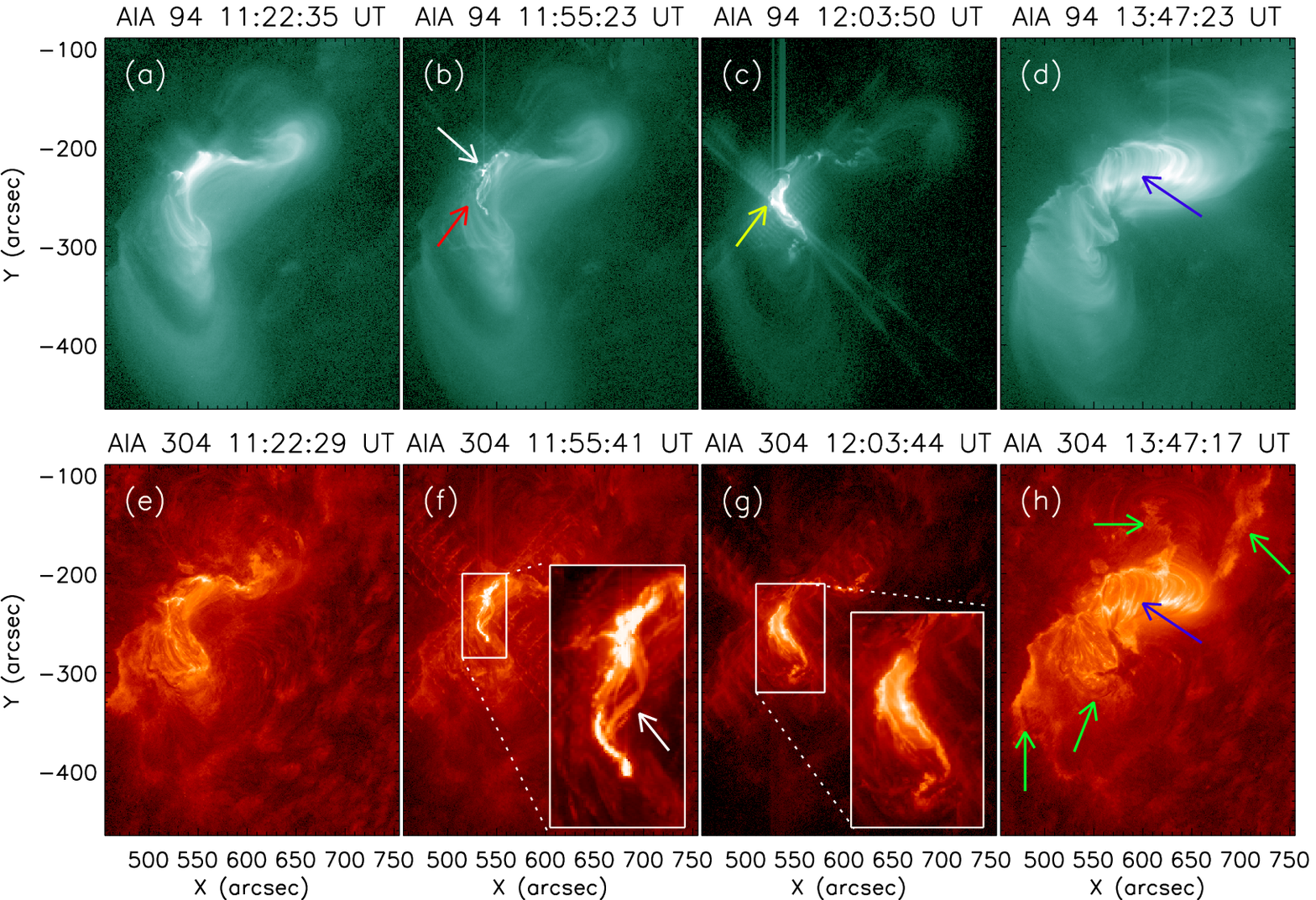} 
\caption{AIA 94~\AA~(panels (a)-(d)) and 304~\AA~(panels (e)-(h)) images showing the evolution of X9.3 flare on 6 September 2017. The red arrow in panel (b) indicates a hot channel. The white arrow in panel (b) indicates the location of the earliest brightening prior to the eruption of the hot channel. The yellow arrow in panel (c) indicates intense thermal emission from the location of the hot channel. The blue arrows in panels (d) and (h) indicate a large-scale post-reconnection arcade encompassing almost the entire span of the active region. The white arrow in the inset of panel (f) indicates the erupting hot filament. The green arrows in panel (h) indicate a set of large-scale flare ribbons during the gradual phase of the X9.3 flare.} \label{fig:X9flare}

\end{center}
\end{figure}

\subsection{Homologous X-class flares on 6 September 2017}
\label{sec2}
%Figure 6.10
%Figure 6.13

\begin{figure}[h]
\begin{center}
\includegraphics[width=0.45\textwidth]{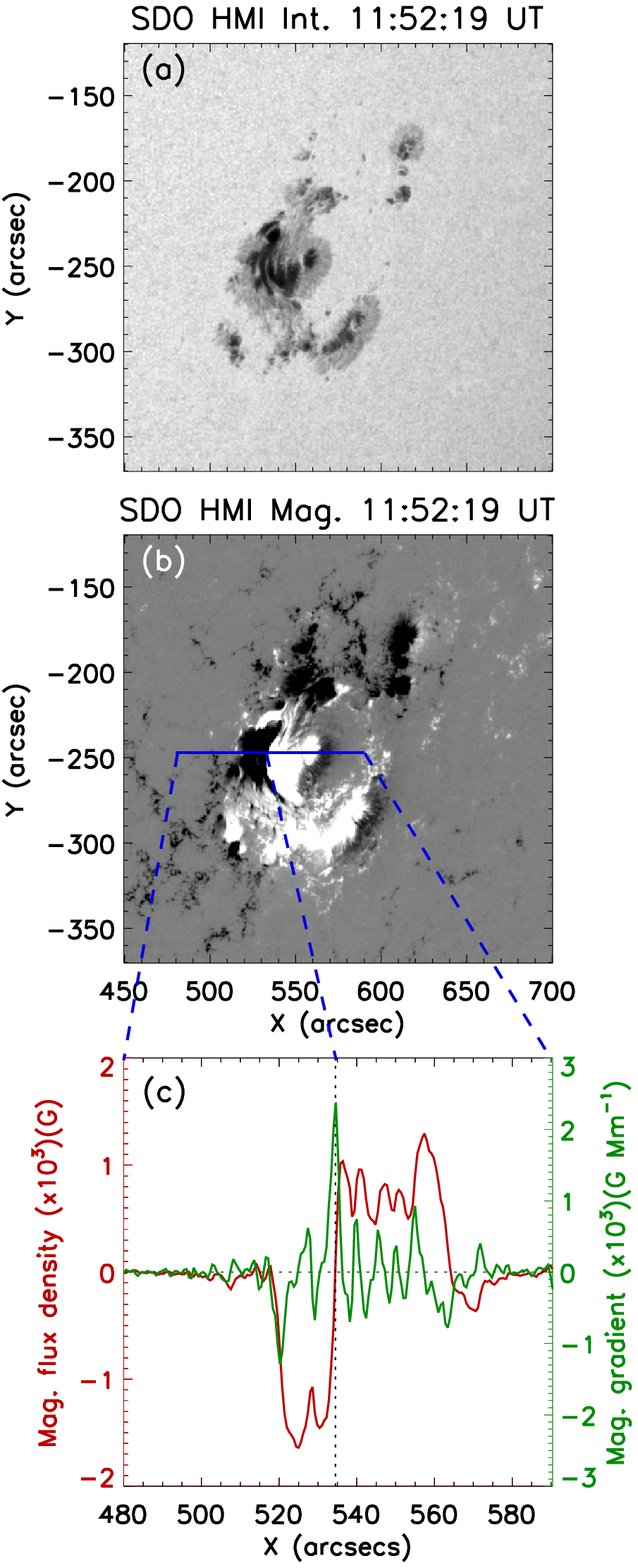} 
\caption{Panel (a): HMI continuum image of the active region NOAA
12763 prior to the onset of the X9.3 flare on 6 September 2017, showing
multiple, fragmented umbrae within a single penumbrae. Panel (b):
co-temporal HMI LOS magnetogram of NOAA 12673. Panel (c): LOS magnetic flux density (red) and magnetic gradient (green) computed along the blue slit in panel (b).} \label{fig_magnetic_structure}
\end{center}
\end{figure}

\begin{figure}[h]
\begin{center}
\includegraphics[width=0.9\textwidth]{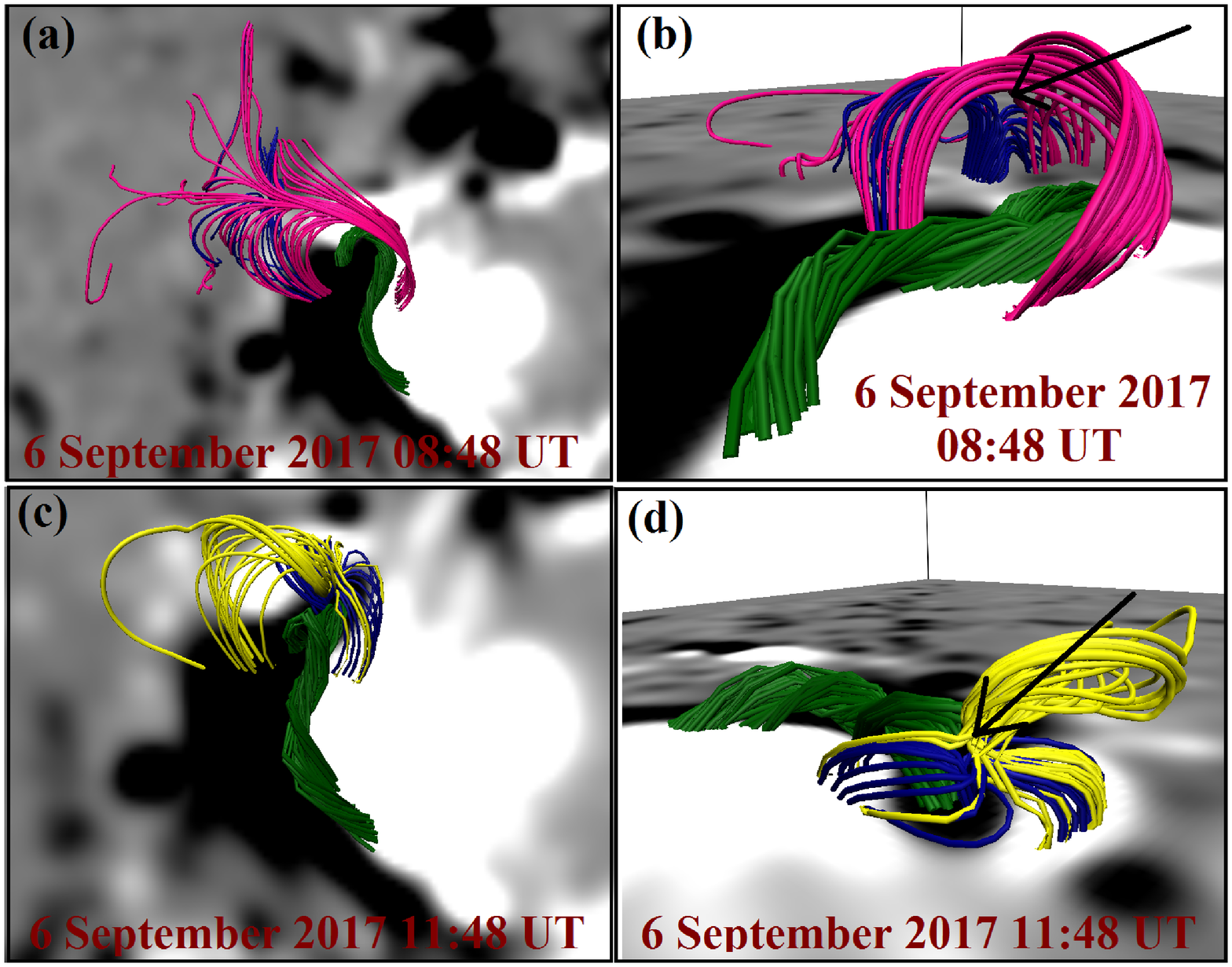} 
\caption{Reconstructed coronal magnetic field configuration by a NLFFF model \citep{Wiegelmann2012} prior to the X2.2 flare (top panel) and X9.3 flare (bottom panel). In all the panels, the green lines represent magnetic flux ropes. The arrows in panels (b) and (d) indicate coronal null points.} \label{fig:NLFFF}
\end{center}
\end{figure}

The first X-class flare from NOAA 12673 (GOES class X2.2) originated on 6 September and within $\approx$3 hours after its initiation,
the AR produced the largest flare of the solar cycle 24, having GOES class X9.3. Figure~\ref{fig:X2flare} represents a few representative 94~\AA~and 304~\AA~images showing the evolution of X2.2 flare. The EUV images prior to the onset of the flare (Figures~\ref{fig:X2flare}(a) and (d)) readily confirms that the coronal configuration associated with the AR took the shape of an impressive `inverse-S' i.e., a prominent coronal sigmoid (the sigmoidal structure is highlighted by the dotted curve in Figure~\ref{fig:X2flare}(a)). Notably, the sigmoid encompassed the entire stretch of the AR and axis of the sigmoid was associated with the PIL between the positive and negative flux regions. The intense emission observed in the 94~\AA~images along the coronal sigmoid is suggestive of the presence of a hot corona channel i.e., an activated flux rope. During the impulsive phase of the flare, this flux rope underwent partial eruption with the formation of chromospheric flare ribbons.

After the X2.2 flare, the sigmoidal structure underwent a gradual expansion
and rise in the intensity of emission (cf. Figures~\ref{fig:X2flare}c and  ~\ref{fig:X9flare}a). Notably, the X9.3 flare initiated with a localized spot-like brightening at the northern footpoint of the hot channel (indicated by the white arrow in Figure~\ref{fig:X9flare}(b)).  Almost simultaneously we observed prominent signatures of filament eruption which initially ascended with the hot channel (indicated by the arrow in the inset of Figure~\ref{fig:X9flare}(f)). Following the eruption, we observed formation of two prominent flare ribbons which were situated close to each other (shown within the inset of Figure~\ref{fig:X9flare}(g). The narrow post-reconnection arcade connecting these flare ribbons is indicated by the yellow arrow in Figure~\ref{fig:X9flare}(c). Importantly, the gradual phase of the flare was characterized by the formation of a large-scale loop arcade and flare ribbons that encompassed the entire stretch of the AR (indicated by the blue arrows in Figures~\ref{fig:X9flare}(d) and (h)). A set of extended flare ribbons at the footpoints of the large-scale loop arcade was identified in the AIA 304 \AA\ images (indicated by the green arrows in Figure~\ref{fig:X9flare}(h)).

\subsection{Complexity and topology of magnetic fields}

To understand the magnetic nature of the $\delta$-spots of NOAA 12673 prior to X-class flares, in Figure~\ref{fig_magnetic_structure}, we plot co-temporal white light image and magnetogram of the AR during its most flaring phase and calculate magnetic gradient across the PIL associated with the bipolar region. The white light image suggests the presence of a few segmented umbrae within the central sunspot area which were surrounded by a single penumbra (Figure~\ref{fig_magnetic_structure}(a) and (b)). Comparison of the white light image and the magnetogram suggests a few of these fragmented umbrae to have
negative polarity and a few to be associated with positive polarity. Our analysis readily suggests that LOS magnetic strength sharply changed from $\approx$1500 G to $\approx$-1000 G within a distance of $\approx$1 arcsec across the PIL. Consequently, the PIL region was associated with the highest magnetic gradient with a value $\approx$ 2.4$\times$10$^{3}$~G~Mm$^{-1}$. This confirms NOAA 12673 to be associated with all the characteristic properties of typical $\delta$-type sunspots prior to the onset of the X-class flares.

In order to understand the coronal magnetic configuration leading to the onset of the two X-class flares, we carried out NLFFF extrapolation at 08:48 UT and 11:48 UT (prior to the onset of the X2.2 and X9.3 flares, respectively). The modeled magnetic field configuration prior to the onset of the X2.2 flare
(Figures~\ref{fig:NLFFF}(a) and (b)) readily indicates the presence of a flux rope extending over the PIL (shown by the green lines) with northern and southern footpoints fixed at negative and positive polarity regions of the central sunspot region, respectively. Notably, the appearance of the hot channel, observed just prior to the onset of the X2.2 flare, was almost precisely the same as the shape of the flux rope (see Figure~\ref{fig:X2flare}(b)). The location next to the northern footpoints of the flux rope was characterized by a complex coronal structure similar to the fan-spine configuration that involved a magnetic null point. The two sets of fan-spine lines are depicted by the blue and pink lines in Figures~\ref{fig:NLFFF}(a) and (b) and the coronal null point is indicated by a black arrow in Figure~\ref{fig:NLFFF}(b).

The overall model magnetic configuration prior to the onset of the X9.3 flare at the core region of NOAA 12673 (Figures~\ref{fig:NLFFF}(c)-(d)) was similar to that corresponding to the X2.2 flare. However, the flux rope prior to the X9.3 flare (shown by the green lines in Figures~\ref{fig:NLFFF}(c)-(d)) appeared to be magnetically more twisted than that of the X2.2 flare. The overall fan-spine configuration prior to the X9.3 flare is shown by the blue and yellow lines in Figures~\ref{fig:NLFFF}(c) and (d). Notably, the northern footpoint of the flux rope itself appeared to be involved in the fan-spine configuration. In Figure~\ref{fig:NLFFF}(d), the null point involved in the fan-spine configuration is indicated by a black arrow. Here we remember that, the X9.3 initiated with a point-like brightening from the northern footpoint location of the hot channel (Figure~\ref{fig:NLFFF}(b)).

\section {Conclusions} 
The AR NOAA 12673 was highly flare productive. Interestingly, the AR emerged on the visible part of the Sun with the simplest magnetic category of solar active regions ($\alpha$-type). However, constant emergence of magnetic flux in the vicinity of the initial sunspot (Figure~\ref{fig:magnetogram}), coupled with photospheric motions enabled the active region to stand out from other active regions by making it one of the most complex active regions with extremely high flaring activity. It is well understood that photospheric motions lead to the storage of non-potential magnetic energy in the active regions making them flare productive \citep{Aulanier2014,Mitra2020SoPh}. Thus, despite having a simplistic initial phase, the active region NOAA 12673 became capable of producing a series of large flares within a few days, thanks to the consistent flux emergence and continuous storage of non-potential energy \citep{Mitra2018,Mitra2020}.

During the occurrence of X-class flares, the complex AR had shown $\delta$-sunspots, which are identified with a complex distribution of sunspot groups in which the umbrae of positive and negative polarities share a common penumbra. Such complex ARs are known to produce powerful flares \citep{Takizawa2015}. The $\delta$-sunspots were concentrated at the central part of the AR where magnetic fields were very strong, and the magnetic field gradient across the PIL was extremely high. The reported eruptive activities in AR 12673, thus, represent, the capability of the AR in the rapid generation and storage of huge amount of excess magnetic energy in the corona. The evolution of free magnetic energy by \cite{Mitra2018} suggest that a large amount of free magnetic energy was already stored in NOAA 12673 before the flaring activities and that the large X-class flares of 6 September 2017 essentially released only a small fraction of it.

Our observations reveal a sigmoid structure \cite[i.e., large, bright AR loops presenting S-shaped morphology; see][]{Joshi2017} before the first X-class
flare that remained intact after the event (Figure~\ref{fig:X2flare}). However, the sigmoidal region completely transformed into the postflare arcade in a sequential manner during the second X-class flare (Figures~\ref{fig:X9flare}). The present observations thus indicate the occurrence of two eruptive X-class flares within a single sigmoid-to-arcade event, which does not seem to be a commonly observed phenomenon. It is noteworthy that the flare ribbons corresponding to the X-class flares were subjected to very little separation, which is rather uncommon in strong eruptive X-class flares. We attribute this small separation between flare ribbons to the fact that footpoints of the flare loops were rooted in $\delta$-sunspot regions of strong magnetic field {\bf B}. Thus, although the loop growth and inner ribbon expansion speed {\bf v} were small, the associated local reconnection rate would have been large enough to produce X-class events.

Both the X-class events not only occurred at the same region but also shared a common initiation process. The trigger happened at the central part of the sigmoid that spatially lay close to the northern edge of the high magnetic gradient area. The detection of preexisting magnetic null in NLFFF extrapolations, near the northern edge of the flux rope structure, has important implications in understanding the triggering mechanism for the subsequent eruptive events \citep[see e.g.,][]{Prasad2020}. The synthesis of observed and modeled coronal loops/field lines provides concrete evidence that the location of the initial energy release is spatially correlated with the site of the magnetic null. Our observations indicate that the triggering reconnections at the magnetic null were capable of destabilizing the flux rope only locally during the first X-class flare because the eruption was relatively minor and the overall sigmoid structure was preserved. This could happen because of the strong field lines lying over the flux rope, as well as firm tying of its southern footpoint. The coronal conditions seem to become conductive for the complete eruption by the time of the second X-class flare, which resulted in the eruptive phenomena at much larger scales.

In summary, flux emergence and photospheric motions stored large amount of non-potential magnetic energy in the active region NOAA 12673. The implications of the non-potential magnetic energy storage can be interpreted in the forms of the complex coronal structures found in the active region i.e., coronal sigmoid, magnetic flux rope, coronal null point etc. Thus, NOAA 12763 can be considered as a test-subject to examine the role of photospheric magnetic activities toward the triggering of powerful solar eruptions.

\begin{acknowledgments}
BJ acknowledges the IAU for providing the travel grant to attend the General Assembly.
\end{acknowledgments}

\section{Bibliography}

%Definition of a few common journal names 
\def\apj{{ApJ}}    
\def\nat{{Nature}}    
\def\jgr{{JGR}}    
\def\apjl{{ApJ Letters}}    
\def\aap{{A\&A}}   
\def\mnras{{MNRAS}}
\def\solphys{{SoPh}}
\def\aj{{AJ}}
\let\mnrasl=\mnras
\def\hia{{Highlights of Astronomy}}

\bibliographystyle{aa}
%\bibliography{biblio.IAUS372}

\end{document}